# Oblivious channels

Michael Langberg
California Institute of Technology
Email: mikel@caltech.edu

*Abstract*—Let $C = \{\mathbf{x}_1, \ldots, \mathbf{x}_N\} \subset \{0,1\}^n$ be an $[n, N]$ binary error correcting code (not necessarily linear). Let $\mathbf{e} \in \{0,1\}^n$ be an error vector. A codeword $\mathbf{x} \in C$ is said to be *disturbed* by the error $\mathbf{e}$ if the closest codeword to $\mathbf{x} \oplus \mathbf{e}$ is no longer $\mathbf{x}$. Let $A_\mathbf{e}$ be the subset of codewords in $C$ that are disturbed by $\mathbf{e}$. In this work we study the size of $A_\mathbf{e}$ in random codes $C$ (*i.e.* codes in which each codeword $\mathbf{x}_i$ is chosen uniformly and independently at random from $\{0,1\}^n$). Using recent results of Vu [Random Structures and Algorithms 20(3)] on the concentration of non-Lipschitz functions, we show that $|A_\mathbf{e}|$ is strongly concentrated for a wide range of values of $N$ and $\|\mathbf{e}\|$.

We apply this result in the study of communication channels we refer to as *oblivious*. Roughly speaking, a channel $W(\mathbf{y}|\mathbf{x})$ is said to be oblivious if the error distribution imposed by the channel is independent of the transmitted codeword $\mathbf{x}$. For example, the well studied Binary Symmetric Channel is an oblivious channel.

In this work, we define oblivious and partially oblivious channels and present lower bounds on their capacity. The oblivious channels we define have connections to Arbitrarily Varying Channels with state constraints.

## I. INTRODUCTION

For a parameter $n$, a general (not necessarily memoryless) binary communication channel $W$ for block length $n$ is a probability distribution over $\{0,1\}^n \times \{0,1\}^n$. Namely $W$ is defined by the conditional probabilities $W(\mathbf{y}|\mathbf{x})$ that $\mathbf{y} \in \{0,1\}^n$ is received when $\mathbf{x} \in \{0,1\}^n$ is transmitted.

An $[n, N]$ binary block code $\mathcal{C}$ is defined by a codebook of $N$ codewords $C = \{\mathbf{x}_1, \ldots \mathbf{x}_N\}$ in $\{0,1\}^n$ corresponding to messages $\{1, \ldots, N\} = [N]$ and a decoder $\phi : \{0,1\}^n \to [N]$. The probability of error for message $i$, when $\mathcal{C}$ is used on a channel $W$ is $e(i) = \sum_{\mathbf{y}:\phi(\mathbf{y}) \neq i} W(\mathbf{y}|\mathbf{x}_i)$.

An $[n, N]$ code $\mathcal{C}$ is said to allow communication at rate $R$ over the channel $W$ with (average) error $\varepsilon > 0$ if $N \geq 2^{Rn}$ and $\bar{e} = \frac{1}{N}\sum_{i=1}^N e(i) < \varepsilon$. An $[n,N]$ code $\mathcal{C}$ is said to allow communication at rate $R$ over a *family* of channels $\mathcal{W}$ with error $\varepsilon$ if for every $W \in \mathcal{W}$ the code $\mathcal{C}$ allows communication at rate $R$ over $W$ with error $\varepsilon$. Rate $R$ is an achievable rate for the family $\mathcal{W}$ if for every $\varepsilon > 0$, $\delta > 0$ and every sufficiently large $n$ there exists an $[n, N]$ code $\mathcal{C}$ such that $\mathcal{C}$ allows communication at rate $\geq R - \delta$ over the family $\mathcal{W}$ with error at most $\varepsilon$[1]. The maximum achievable rate is called the capacity of the family $\mathcal{W}$, and is denoted by $\mathcal{C}(\mathcal{W})$.

When considering the capacity of a family of channels $\mathcal{W}$, one must address the design of error correcting codes which allow communication under the uncertainty of which channel $W$ is actually used from the family $\mathcal{W}$. Intuitively, this corresponds to the design of codes which allow communication in an *adversarial* jamming model in which an entity $Z$ controlling the channel is assumed to act maliciously within the limits of $\mathcal{W}$. We will adapt this interpretation in the discussions throughout this work.

### A. This work

Several families of channels have been studied over the last few decades (for a nice survey on communication under channel uncertainty see [10]). For a constant $p \in (0, 1/2)$ a $p$-channel $W$ is a channel for which $W(\mathbf{y}|\mathbf{x}) = 0$ if the Hamming[2] distance between $\mathbf{x}$ and $\mathbf{y}$ is greater than $pn$. In words, a $p$-channel can only change at most $pn$ entries of $\mathbf{x}$. The parameter $p$ may be viewed as the amount of *power* that can be used by the channel when imposing an error. In this work we study the capacity of various families of binary $p$-channels.

A natural starting point is the extensively studied family $\mathcal{W}_p$ of all binary $p$-channels. The capacity of $\mathcal{W}_p$ is a long standing open problem. There is a strong connection between codes $\mathcal{C}$ that allow communication over $\mathcal{W}_p$ and the minimal distance of $\mathcal{C}$. Namely, $\mathcal{C}(\mathcal{W}_p)$ equals the maximum (asymptotic) rate of $[n, N]$ block codes with minimum distance greater than $2pn$ (a detailed proof appears in the Appendix). The latter rate is not known. It is known that this rate is bounded away from $1 - H(p)$ (*e.g.* [2], [13], [15]), while the currently best known lower bound stands on $1 - H(2p)$ (Gilbert-Varshamov [7], [16]).

We will not study the capacity of $\mathcal{W}_p$, rather we turn to study certain subfamilies $\mathcal{W} \subseteq \mathcal{W}_p$. Consider the adversarial model discussed above, in which an adversarial entity $Z$ may choose which channel $W \in \mathcal{W}$ to use based on the code $\mathcal{C}$ shared by the sender and receiver. In the case of communication over $\mathcal{W}_p$ this adversarial entity $Z$ is very powerful as it can choose any $p$-channel $W$ and *tailor* the error it imposes to fit not only the code $\mathcal{C}$ in use but also the codeword $\mathbf{x}$ transmitted. Indeed, $Z$ can use a channel $W(\mathbf{y}|\mathbf{x}) \in \mathcal{W}_p$ in which the error distribution imposed by the channel strongly depends on the transmitted codeword $\mathbf{x}$.

In this work we study scenarios in which $Z$ is limited in its dependence on $\mathbf{x}$. Specifically, we study the scenario in which the error imposed by $Z$ is *oblivious* or partly oblivious of the codeword $\mathbf{x}$ transmitted. For example, if $Z$ always imposes exactly the same distribution over errors, no matter which codeword $\mathbf{x}$ is sent, then $Z$ is said to be completely oblivious of $\mathbf{x}$. A well studied oblivious channel is the Binary Symmetric Channel with cross over probability $p$. We denote this channel by $W_{BSC_p}$. Indeed, no matter which codeword $\mathbf{x}$ is transmitted the error imposed by $W_{BSC_p}$ follows the same distribution. In this work we define and study families of channels with varying degrees of obliviousness.

---

[1]In the study of communication over families of channels it is also common to address the *maximum* error $e = \max_i e(i)$ instead of $\bar{e}$; and the rate achievable when using a distribution over codes (random coding) instead of a deterministic code $\mathcal{C}$ as above. These models are briefly addressed in the Appendix.

[2]Let $\mathbf{x} = x_1 x_2 \ldots x_n$ be an element in $\{0,1\}^n$. The Hamming weight $\|\mathbf{x}\|$ is defined to be the number of positions $i$ in which $x_i \neq 0$.

*B. Oblivious channels*

We start by giving a slightly different (but equivalent) definition of a binary channel $W$. Instead of defining $W$ in terms of the conditional probabilities $W(\mathbf{y}|\mathbf{x})$, one may define $W$ in terms of the conditional probabilities $W(\mathbf{e}|\mathbf{x})$; where $\mathbf{e} \in \{0,1\}^n$ is the error imposed by the channel $W$. Specifically, in this setting $\mathbf{y} = \mathbf{x} \oplus \mathbf{e}$. For example, by our definitions, a $p$-channel $W$ is a channel for which $W(\mathbf{e}|\mathbf{x}) = 0$ for every $\mathbf{e}$ of Hamming weight above $pn$. Let $\Pi$ be the set of distributions over errors $\mathbf{e} \in \{0,1\}^n$. In this setting, a channel $W$ may be viewed as a function from $\mathbf{x} \in \{0,1\}^n$ to the set $\Pi$. Now we are ready to define $\gamma$-oblivious channels for $\gamma \in [0,1]$.

Roughly speaking, a channel $W : \{0,1\}^n \to \Pi$ is said to be oblivious if it is a constant function. In this case we will say that $W$ is 1-oblivious. The obliviousness of a channel is determined by the size of its image. Namely, channels $W$ with image size at most $2^n$ will be referred to as 0-oblivious channels (thus any channel is 0-oblivious). For $\gamma \in [0,1]$ channels with image size at most $2^{(1-\gamma)n}$ will be referred to as $\gamma$-oblivious.

*Definition 1.1:* A channel $W$ with block length $n$ is $\gamma$-oblivious if there is a $2^{(1-\gamma)n}$ sized family of distributions $\pi = \{\pi_1, \ldots, \pi_{2^{(1-\gamma)n}}\} \subset \Pi$, such that for every $\mathbf{x} \in \{0,1\}^n$ the marginal distribution $W(\cdot|\mathbf{x})$ over $\mathbf{e}$ is in the set $\pi$. A family of channels $\mathcal{W}$ is $\gamma$-oblivious if for each $W \in \mathcal{W}$, $W$ is $\gamma$-oblivious.

For example, the Binary Symmetric Channel is 1-oblivious, as $W_{BSC_p}(\mathbf{e}|\mathbf{x})$ is completely independent of $\mathbf{x}$; and the family $\mathcal{W}_p$ is 0-oblivious (and not $\gamma$-oblivious for any $\gamma > 0$). Let $\mathcal{W}_{p,\gamma}$ be the family of all $p$-channels that are $\gamma$-oblivious. In this work we study the capacity of $\mathcal{W}_{p,\gamma}$ for various values of $p$ and $\gamma$. The main result of this work can be summarized in the following Theorem.

*Theorem 1:* For any $p \in [0, 1/2]$ and any $\gamma \in \left(\frac{2+H(p)}{3}, 1\right]$

$$\mathcal{C}(\mathcal{W}_{p,\gamma}) \geq \gamma - H(p).$$

A few remarks are in place. It is not hard to verify (detailed proof appears in the Appendix) that for $\gamma = 1$, Theorem 1 is tight. Namely, $\mathcal{C}(\mathcal{W}_{p,1}) = 1 - H(p)$ (the capacity of $W_{BSC_p}$ [14]), this follows from the fact that $W_{BSC_p}$ is a 1-oblivious channel which in *essence*[3] is also a $p$-channel. It also holds that $\mathcal{C}(\mathcal{W}_{p,\gamma}) \geq \mathcal{C}(\mathcal{W}_p) \geq 1 - H(2p)$. A simple calculation shows that $1 - H(2p)$ may be above the bound of Theorem 1 only for very small $p \leq 0.07$ and $\gamma \in \left[\frac{2+H(p)}{3}, 1 - H(2p) + H(p)\right)$.

---
[3]Notice that $W_{BSC_p}$ is not a $p$-channel, however the error it imposes is expected to be of Hamming weight $pn$.

The study of $\mathcal{C}(\mathcal{W}_{p,\gamma})$ arises when considering communication in an adversarial jamming model in which the jammer $Z$ is limited in resources. Primarily, we restrict the jammer to flip at most a $p$-fraction of the bits transmitted, which corresponds to a power constraint imposed on $Z$. In addition, we limit the jammer's view of the transmitted codeword. This is obtained by forcing the jammer to use a channel $W$ which can not properly differentiate between different codewords $\mathbf{x}$. Namely, by restricting $W$ to impose its error based on only a small number of possible error distributions, it must be the case that the exact same distribution is used on large portions of codewords.

An alternative (but problematic) definition to $\gamma$-oblivious channels $W$ that may come in mind is one in which we restrict $\max_X I(X;Z)$ to be at most $(1-\gamma)n$. Here $X$ represents any distribution over codewords transmitted and $Z$ denotes the error imposed by the channel. The random variables $X$ and $Z$ are jointly distributed according to $W(\mathbf{e}|\mathbf{x})$. There are various connections between the suggested definition and the original one given in Definition 1.1. However, they are not equivalent, and roughly speaking, the suggested definition implies a discontinuous capacity function at the point $\gamma = 1$. A detailed discussion appears in the Appendix.

*C. Previous results and connection to AVC's*

To the best of our knowledge, $\gamma$-oblivious $p$-channels for general $\gamma \in [0,1]$ have not been addressed in the past. For the special case $\gamma = 1$, as we state shortly, there is a strong connection between 1 oblivious $p$-channels and so called arbitrarily varying channels (AVC) with state constraints.

A (discrete memoryless) arbitrarily varying channel [3] of block length $n$ is a family of channels $\mathcal{W}$ defined by a set of *states* $S$ and a set of channels $\mathcal{S} = \{W_s(y|x) | s \in S\}$ of block length 1 (in the binary case $x$ and $y$ are in $\{0,1\}$). Specifically, the family $\mathcal{W}_\mathcal{S}$ that corresponds to $\mathcal{S}$ consists of the channels $\{W_\mathbf{s} | \mathbf{s} \in S^n\}$ defined by $W_\mathbf{s}(\mathbf{y}|\mathbf{x}) = \Pi_{i=1}^n W_{s_i}(y_i|x_i)$. In the above, $\mathbf{x} = x_1, \ldots, x_n$; $\mathbf{y} = y_1, \ldots, y_n$; and $\mathbf{s} = s_1, \ldots, s_n$. If we associate with each state $s \in S$ a cost $\ell(s)$, an AVC family with state constraint $p$ is the family of channels $W_\mathbf{s} \in \mathcal{W}_\mathcal{S}$ for which $\frac{1}{n} \sum_{i=1}^n \ell(s_i) \leq p$.

Consider the binary 1-block channels $W_0$ and $W_1$ defined by $W_s(y|x) = 1$ iff $(x+s = y)$ modulo 2. Let $\ell(s) = s$ for $s \in \{0,1\}$. Let $\mathcal{W}^*$ denote the AVC family defined by $W_0$ and $W_1$ with state constraint $p$. The families $\mathcal{W}_{p,1}$ and $\mathcal{W}^*$ are closely related and it holds that $\mathcal{C}(\mathcal{W}_{p,1}) = \mathcal{C}(\mathcal{W}^*)$.

The capacity of AVC with state (and also input) constraints was studied extensively in the works of Csiszár and Narayan [4], [5]. Using proof techniques that build strongly upon the *method of types*, Csiszár and Narayan show that the capacity of $\mathcal{C}(\mathcal{W}^*)$ is $1 - H(p)$. Thus, proving Theorem 1 for the case $\gamma = 1$. The proof presented in this work differs substantially from the proofs of Csiszár and Narayan. Namely, our proof technique is combinatorial in nature and is based on a relatively new "strong concentration inequality" of [17]. This inequality and its application in the context of coding theory may be of independent interest.

For $\gamma < 1$, $\gamma$-oblivious channels were not defined or discussed in [4], [5]. However, a careful examination of their proof techniques yields an implicit bound on the capacity of $\mathcal{C}(\mathcal{W}_{p,\gamma})$ for large values of $\gamma$. Namely, it can be shown using the proof techniques that appear in [4] that $\mathcal{C}(\mathcal{W}_{p,\gamma}) \geq 1 - H(p) - 30(1-\gamma)$. For comparison using our proof techniques we show that $\mathcal{C}(\mathcal{W}_{p,\gamma}) \geq 1 - H(p) - (1-\gamma)$.

### D. Proof Techniques, random codes, and list decodable codes

To prove the lower bound of Theorem 1 we need to show the existence of high rate codes $\mathcal{C}$ which enable communication over $\gamma$-oblivious $p$-channels. We first note that no linear code will suffice. Roughly speaking, this follows from the fact that each codeword $\mathbf{x}_i$ in a linear code $\mathcal{C}$ has exactly the same "neighborhood structure". Thus, when a linear code is used, the problem of communicating over the oblivious or partially oblivious families $\mathcal{W}_{p,\gamma}$ is equivalent to that of communication over $\mathcal{W}_p$ (a detailed proof appears in the Appendix). We thus turn to study codes which are not linear. A natural candidate is a code $\mathcal{C}$ in which the codewords $C = \{\mathbf{x}_1, \ldots, \mathbf{x}_N\}$ are chosen completely at random, (*i.e.* a code in which each codeword is chosen uniformly and independently from $\{0,1\}^n$, and $\phi$ is the Nearest Neighbor decoder. Let $\mathbf{e} \in \{0,1\}^n$ be an error vector of Hamming weight at most $pn$. A codeword $\mathbf{x}$ is said to be *disturbed* by the error $\mathbf{e}$ if the closest codeword to $\mathbf{x} \oplus \mathbf{e}$ is no longer $\mathbf{x}$. Let $A_\mathbf{e} = A_\mathbf{e}(C)$ be the subset of codewords $\mathbf{x}$ in $C$ that are disturbed by $\mathbf{e}$. In Section II we show that $\mathcal{C}$ enables communication over all $\gamma$-oblivious $p$-channels if for every error $\mathbf{e}$ of Hamming weight at most $pn$ the size of $A_\mathbf{e}$ is relatively *small*.

Hence, it suffices to analyze the size of $A_\mathbf{e}$ over random codebooks $C$. Specifically we are interested in showing that with positive probability $A_\mathbf{e}$ is small for every error $\mathbf{e}$ of weight at most $pn$. Let $R = \gamma - H(p)$. It is straightforward to verify that for a fixed error $\mathbf{e}$, the expected size of $A_\mathbf{e}$ taken over random $C = \{\mathbf{x}_1, \ldots, \mathbf{x}_{\lfloor 2^{Rn} \rfloor}\}$ is relatively small. Hence it is left to show that with high probability $|A_\mathbf{e}|$ does not deviate significantly from its expectation. Indeed if this is the case, a simple union bound will imply our assertion.

Strong concentration (or large deviation) inequalities have been extensively studied. The usual way to prove such inequalities is via the Azuma or Talagrand inequalities (*e.g.* [1]). These inequalities work very well when the random variable at hand has a small *Lipschitz coefficient*. In our case the Lipschitz coefficient of $|A_\mathbf{e}|$ is defined by the maximum of $||A_\mathbf{e}(C)| - |A_\mathbf{e}(C')||$ where $C$ and $C'$ are two codebooks which differ only in a single codeword. It is not hard to verify that the Lipschitz coefficient of $|A_\mathbf{e}|$ may be very large. However, we show that for most pairs $C$ and $C'$ as above, the difference $||A_\mathbf{e}(C)| - |A_\mathbf{e}(C')||$ is relatively small and is bounded by the *list decoding* quality of $C$ (the maximal number of codewords in $C$ which are included in a Hamming ball of radius $pn$). With this in mind, we are able to use a recent result of Vu [17] on the concentration of random variables with large *worst case* Lipschitz coefficients but small *average* case coefficients. The application of the framework suggested in [17] to our random variable $|A_\mathbf{e}|$ is somewhat involved and can be viewed as the main technical contribution of this paper.

There are other proof techniques which are common in the study of probabilistic combinatorics. For example, so called "correlation inequalities" (*e.g.* [1]) are often used to analyze the probability of the intersection of many events. We would like to note that such inequalities may also be used to study the problem phrased above, however they only yield results for small values of $p$ that satisfy $H(p) \leq \frac{1}{2}$, as in this case the number of events considered is relatively small.

*Definition 1.2:* Let $\Omega[n, N]$ be the distribution over $[n, N]$ codebooks $C = \{\mathbf{x}_1, \ldots, \mathbf{x}_N\}$ in which each codeword in $C$ is chosen uniformly and independently from $\{0,1\}^n$.

*Definition 1.3:* For $\mathbf{x} \in \{0,1\}^n$ and integer $r$, let $\mathcal{B}(r, \mathbf{x})$ be the Hamming ball of radius $r$ centered at $\mathbf{x}$.

*Definition 1.4:* For a given codebook $C = \{\mathbf{x}_1, \ldots, \mathbf{x}_N\}$, and error $\mathbf{e} \in \{0,1\}^n$, let $A_\mathbf{e}(C) = \{\mathbf{x}_i | \exists j \neq i \text{ s.t. } \mathbf{x}_j \in \mathcal{B}(\|\mathbf{e}\|, \mathbf{x}_i \oplus \mathbf{e})\}$. When the reference codebook $C$ is clear we will denote $A_\mathbf{e}(C)$ by $A_\mathbf{e}$.

*Theorem 2:* Let $p \in [0, 1/2)$. Let $\gamma \in \left(\frac{2+H(p)}{3}, 1\right]$. Let $\delta > 0$ be any sufficiently small constant. Let $R = \gamma - H(p) - \delta$. Let $n$ be sufficiently large. Let $\mathbf{e}$ be any error vector in $\{0,1\}^n$ of Hamming weight at most $pn$. Then $\Pr[|A_\mathbf{e}| - \mathbb{E}(|A_\mathbf{e}|) > 2^{(H(p)+2R-1)n}] \leq 2^{-2n}$. Here the probability is over $\Omega[n, \lfloor 2^{Rn} \rfloor]$.

The remainder of this work is organized as follows. In Section II we present some preliminaries on the distribution $\Omega[n, N]$ and on oblivious channels. In Section III we present the proof of Theorem 2 (which will imply Theorem 1).

## II. PRELIMINARIES

For any integer $i$, let $[i]$ denote the set $\{1, 2, \ldots, i\}$. Let $H(x) = -x \log_2 x - (1-x) \log_2(1-x)$ be the standard (binary) entropy function. For a codebook $C = \{\mathbf{x}_1, \ldots, \mathbf{x}_N\}$, the corresponding *Nearest Neighbor* decoder is the decoder $\phi$ which on input $\mathbf{y} \in \{0,1\}^n$, returns the index of the closest codeword $\mathbf{x}_i$ in $C$ to $\mathbf{y}$. For uniqueness, we will assume ties are broken by the natural lexicographic ordering on $\{0,1\}^n$. To simplify notation, for any $R \in [0,1]$ and integer $n$, we assume throughout that $2^{Rn}$ is integer.

*Definition 2.1 (List decodability):* An $[n, N]$ binary codebook $C$ is said to be $[\ell, p]$ list decodable iff $|C \cap \mathcal{B}(pn, \mathbf{y})| \leq \ell$ for any $\mathbf{y} \in \{0,1\}^n$.

We first analyze the list decoding properties of random codes. The lemma that follows has appeared in various forms in the past (*e.g.* [6], [18]). Full proof is given in the Appendix.

*Lemma 2.1:* Let $R \leq 1 - H(p)$. Let $n$ be sufficiently large. Let $C$ be a random codebook in $\Omega[n, 2^{Rn}]$. With probability $1 - 2^{-n^2}$, $C$ is $[12n^2, p]$ list decodable.

Let $\mathbf{e}$ be an error in $\{0,1\}^n$. Recall the definition of $A_\mathbf{e}(C)$ from Definition 1.4. We now define an alternative sufficient condition for a code $C$ to allow communication over $\gamma$-oblivious $p$-channels. We will use this sufficient condition throughout our work.

*Lemma 2.2:* An $[n, 2^{Rn}]$ codebook $C$ with the Nearest Neighbor decoder $\phi$ allows communication over $\mathcal{W}_{p,\gamma}$ within error $\varepsilon$ if for every error $\mathbf{e} \in \mathcal{B}(pn, \mathbf{0})$ it is the case that $|A_\mathbf{e}|$ is at most $\varepsilon 2^{(R-(1-\gamma))n}$.

*Proof:* Let $C = \{\mathbf{x}_1, \ldots, \mathbf{x}_{2^{Rn}}\}$ be a codebook in which for every error $\mathbf{e} \in \mathcal{B}(pn, \mathbf{0})$ it is the case that $|A_\mathbf{e}|$ is at most $\varepsilon 2^{(R-(1-\gamma))n}$. Let $\phi$ be the Nearest Neighbor decoder. Let $N = 2^{Rn}$. Let $W$ be a channel in $\mathcal{W}_{p,\gamma}$. By Definition 1.1 and the fact that $W$ is a $p$-channel there exists a family of distributions $\pi = \{\pi_1, \ldots, \pi_{2^{(1-\gamma)n}}\}$ over $\mathcal{B}(pn, \mathbf{0})$ of size $2^{(1-\gamma)n}$ such that for every $\mathbf{x} \in \{0,1\}^n$ the marginal distribution $W(\cdot|\mathbf{x})$ over $\mathbf{e}$ is in the set $\pi$. For $i \in [2^{(1-\gamma)n}]$ let $X_i$ be the subset of codewords $\mathbf{x}$ in $C$ for which $W(\cdot|\mathbf{x}) = \pi_i(\cdot)$. We show that $C$ allows communication over $W$ with error at most $\varepsilon$.

$$\begin{aligned}
\bar{e} &= \frac{1}{N} \sum_{i=1}^{N} \sum_{\mathbf{e}:\phi(\mathbf{e}\oplus\mathbf{x}_i)\neq i} W(\mathbf{e}|\mathbf{x}_i) \leq \frac{1}{N} \sum_{\mathbf{e}\in\mathcal{B}(pn,\mathbf{0})} \sum_{\mathbf{x}\in A_\mathbf{e}} W(\mathbf{e}|\mathbf{x}) \\
&= \frac{1}{N} \sum_{i=1}^{2^{(1-\gamma)n}} \sum_{\mathbf{e}\in\mathcal{B}(pn,\mathbf{0})} \sum_{\mathbf{x}\in A_\mathbf{e}\cap X_i} \pi_i(\mathbf{e}) \\
&\leq \frac{1}{N} \sum_{i=1}^{2^{(1-\gamma)n}} \sum_{\mathbf{e}\in\mathcal{B}(pn,\mathbf{0})} \pi_i(\mathbf{e})|A_\mathbf{e}| = \varepsilon
\end{aligned}$$

∎

## III. PROOF OF THEOREM 2

In what follows we prove Theorem 2. We use the notation outlined in the statement of Theorem 2. Let $N = 2^{Rn}$, and $M = 2^n$. We occasionally identify codewords in $C$ with their corresponding messages in $[N]$ and elements in $\{0,1\}^n$ with integers in $[M]$. We first analyze the expected size of $A_\mathbf{e}$ over random codebooks $(\Omega[n, 2^{Rn}])$. For technical reasons, throughout this section we treat codebooks $C$ as *ordered* sets $\langle \mathbf{x}_1, \ldots, \mathbf{x}_N \rangle$ (instead of unordered sets). Accordingly, we change the definition of $\Omega[n, 2^{Rn}]$ to be the uniform distribution over ordered codebooks.

*Lemma 3.1:* $\mathbb{E}[|A_\mathbf{e}|] \leq 2^{(H(p)+2R-1)n}$.

*Proof:* For $i \in [N]$ let $A_\mathbf{e}^i$ be the indicator of the event "$\mathbf{x}_i \in A_\mathbf{e}$". Hence, $\mathbb{E}[|A_\mathbf{e}|] = \sum_i \mathbb{E}[A_\mathbf{e}^i]$. We turn to analyze $\mathbb{E}[A_\mathbf{e}^i]$ for any given $i$. This value is exactly the probability that the ball centered at $\mathbf{x}_i \oplus \mathbf{e}$ of radius $\|\mathbf{e}\|$ includes an additional codeword $\mathbf{x}_j$. For a fixed $j \neq i$, this probability is at most $2^{H(p)n}/2^n$. Here we use the fact that the size of a Hamming ball of radius $pn$ is bounded by $2^{H(p)n}$ [12]. Thus, using the union bound on all $j \neq i \in [N]$, the value of $\mathbb{E}[A_\mathbf{e}^i]$ is bounded by $2^{H(p)n+Rn}/2^n$. This in turn implies that $\mathbb{E}[|A_\mathbf{e}|] \leq 2^{(H(p)+2R-1)n}$. ∎

We now turn to show that the size of $A_\mathbf{e}$ is strongly concentrated. The Lipschitz coefficients of $A_\mathbf{e}$ can be described by the following function $\Delta$. For any $[n, N]$ codebook $C = \langle \mathbf{x}_1, \ldots, \mathbf{x}_N \rangle$, any $i \in [N]$, and any $\mathbf{x} \in \{0,1\}^n$ let $\Delta(i, \mathbf{x}, C) = |\mathbb{E}(|A_\mathbf{e}| : \mathbf{x}_1, \ldots, \mathbf{x}_{i-1}, \mathbf{x}_i = \mathbf{x}) - \mathbb{E}(|A_\mathbf{e}| : \mathbf{x}_1, \ldots, \mathbf{x}_{i-1})|$.

The expectation above is over $\Omega[n, N]$. Given a small global upper bound on the value of $\Delta$ one can prove the tight concentration of $|A_\mathbf{e}|$ using Azuma's inequality. However, it is not hard to verify that $\Delta$ does not have a small global bound in the case under study ($\Delta$ can be as large as a constant fraction of $N$). Nevertheless, as we will show, the value of $\Delta$ is small *on average* and lends itself to the framework outlined in [17], implying the desired concentration. Details follow.

Let $\ell = 12n^2$ be the list decoding parameter from Lemma 2.1. Using a slight change of notation which fits our needs, in Lemma 3.1 of [17] it is shown that:

*Lemma 3.2 (**Lemma 3.1 [17]**):* Let

$$p_1 = \sum_{i=1}^{N} \Pr[\exists \mathbf{x} \in \{0,1\}^n \text{ s.t. } \Delta(i, \mathbf{x}, C) \geq \ell + 3],$$

$$p_2 = \Pr\left[\sum_{i=1}^{N} \sum_{\mathbf{x}\in\{0,1\}^n} \frac{1}{M} \Delta(i, \mathbf{x}, C) \geq N(\ell+3)\right]$$

For any $\lambda \leq 4N$

$$\Pr\left[||A_\mathbf{e}| - \mathbb{E}(|A_\mathbf{e}|)| \geq \sqrt{\lambda N (\ell+3)^2}\right] \leq 2e^{-\lambda/4} + p_1 + p_2.$$

All probabilities and expectation are over $\Omega[n, N]$.

Thus, to use the concentration results of [17] we must bound $p_1$ and $p_2$ defined above. We start by computing the value of $\Delta(i, \mathbf{x}, C)$ for a given $[n, N]$ codebook $C = \langle \mathbf{x}_1, \ldots, \mathbf{x}_N \rangle$. We use the following definitions. For a codebook $C$ and an index $i \in [N]$ let $C|_i$ be the set of ordered $[n, N]$ codebooks that agree with $C$ on the first $i$ codewords, namely a codebook $C' = \langle \mathbf{x}'_1, \ldots, \mathbf{x}'_N \rangle \in C|_i$ iff $\forall j \leq i$ it holds that $\mathbf{x}_j = \mathbf{x}'_j$. For an $[n, N]$ codebook $C = \langle \mathbf{x}_1, \ldots, \mathbf{x}_N \rangle$, an index $i \in [N]$, and $\mathbf{x} \in \{0,1\}^n$ let $C(i, \mathbf{x})$ be the codebook that agrees with $C$ on all but the $i$'th codeword, and on the $i$'th codeword equals $\mathbf{x}$. Recall that $\ell = 12n^2$. An $[n, N]$ codebook $C$ is said to be *typical* if it is $[\ell, p]$ list decodable (the rest are referred to as codebooks which are not typical). Denote the set of typical $[n, N]$ codebooks by $\mathcal{T}$ and codebooks which are not typical by $\mathcal{T}^c$. By Lemma 2.1, at most a fraction of $2^{-n^2}$ (ordered) codebooks are not typical (*i.e.* $|\mathcal{T}^c| \leq M^N 2^{-n^2}$). Notice that the size of $C|_{i-1}$ is $M^{N-i+1}$. Our definitions now imply that

$$\Delta(i, \mathbf{x}, C) = \sum_{C'\in C|_{i-1}} \frac{||A_\mathbf{e}(C'(i,\mathbf{x}))| - |A_\mathbf{e}(C')||}{M^{N-i+1}}$$

We now analyze the value of $||A_\mathbf{e}(C(i, \mathbf{x}))| - |A_\mathbf{e}(C)||$ and show its connection to the list decoding properties of $C$.

*Lemma 3.3:* If an $[n, N]$ codebook $C$ is typical then $||A_\mathbf{e}(C(i, \mathbf{x}))| - |A_\mathbf{e}(C)|| \leq \ell + 2$. If $C$ is not typical then $||A_\mathbf{e}(C(i, \mathbf{x}))| - |A_\mathbf{e}(C)|| \leq N$.

*Proof:* For the first part of the lemma notice that if $C$ is $[\ell, p]$ list decodable then $C(i, \mathbf{x})$ is $[\ell+1, p]$ list decodable. Recall that a codeword $\mathbf{x}_j$ of $C$ is said to be disturbed by the error $\mathbf{e}$ if $\mathbf{x}_j \in A_\mathbf{e}(C)$. The value of $|A_\mathbf{e}(C(i, \mathbf{x}))| - |A_\mathbf{e}(C)|$ is bounded by the maximum number of codewords $\mathbf{x}_j$ disturbed by the error $\mathbf{e}$ *exclusively* due to the change of $\mathbf{x}_i$. Namely, this value is bounded by $|\{j \mid \|\mathbf{x}_i \oplus \mathbf{x}_j \oplus \mathbf{e}\| \leq \|\mathbf{e}\|\}| + 1$ (an additional value of 1 is added for the case that $\mathbf{x}_i$ may be disturbed by $\mathbf{e}$). This in turn is at most $|\{j|\mathbf{x}_j \in \mathcal{B}(pn, \mathbf{x}_i \oplus \mathbf{e})\}| + 1 \leq \ell + 2$. An analogous analysis can be done for

$|A_\mathbf{e}(C)| - |A_\mathbf{e}(C(i,\mathbf{x}))|$. The second part of the lemma follows from the fact that $|A_\mathbf{e}|$ is bounded by $N$. ∎

*Corollary 3.1:* Let $\Gamma$ be the size of $C|_{i-1} \setminus \mathcal{T}$. $\Delta(i,\mathbf{x},C) \leq M^{-(N-i)}\Gamma + \ell + 2$.

We now analyze $p_1$ and $p_2$ of Lemma 3.2.

*Lemma 3.4:* $p_1 \leq 2^{-n^2} MN$.

*Proof:* Let $i \in [N]$. We first note that Corollary 3.1 implies that $\Delta(i,\mathbf{x},C) \geq \ell + 3$ only if the size of $C|_{i-1} \setminus \mathcal{T}$ is at least $M^{N-i}$. Moreover, by our definitions $|\mathcal{T}^c| \leq M^N 2^{-n^2}$ (recall that $\mathcal{T}^c$ is the set of codebooks which are not typical). We now use these facts to prove our assertion.

Notice that for two codebooks $C$ and $C'$ the sets $C|_{i-1}$ and $C'|_{i-1}$ are either equal or disjoint. We partition the set of codebooks in $\Omega[n,N]$ to $M^{i-1}$ disjoint subsets of the form $C|_{i-1}$. Denote these subsets by $\Omega_1, \ldots, \Omega_{M^{i-1}}$. Let $\alpha$ denote the number of these subsets that satisfy $|\Omega_j \setminus \mathcal{T}| \geq M^{N-i}$. As these sets are disjoint and $|\mathcal{T}^c| \leq M^N 2^{-n^2}$; $\alpha$ is at most $M^i 2^{-n^2}$. Finally, for a given $i$, $\Pr[\exists \mathbf{x} \in \{0,1\}^n \text{ s.t. } \Delta(i,\mathbf{x},C) \geq \ell + 3] \leq M^{-(i-1)} \alpha \leq M^{-(i-1)} M^i 2^{-n^2} \leq 2^{-n^2} M$. ∎

*Lemma 3.5:* $p_2 \leq M 2^{-n^2}$.

*Proof:* Consider a codebook $C$, and the event "$\sum_{i=1}^N \sum_{\mathbf{x} \in \{0,1\}^n} \frac{1}{M} \Delta(i,\mathbf{x},C) \geq N(\ell+3)$". This event is included in the event $\sum_{i=1}^N \max_{\mathbf{x} \in \{0,1\}^n} \Delta(i,\mathbf{x},C) \geq N(\ell+3)$. The above event holds only if the size of the set $\{i | \max_{\mathbf{x} \in \{0,1\}^n} \Delta(i,\mathbf{x},C) \geq \ell+3\}$ is greater than 1. We call a codebook $C$ bad if $\{i | \max_{\mathbf{x} \in \{0,1\}^n} \Delta(i,\mathbf{x},C) \geq \ell+3\} \neq \phi$. For each bad $C$ let $d(C) = i - 1$ where $i$ is the minimum integer in $\{i | \max_{\mathbf{x} \in \{0,1\}^n} \Delta(i,\mathbf{x},C) \geq \ell+3\}$.

We now show that the number of bad $C$'s is less than $M^{N+1} 2^{-n^2}$, which concludes our assertion. Consider the set $\mathbb{B} = \mathbb{B}_1$ of bad codebooks $C$ (the indexing of $\mathbb{B}$ will be clear shortly). Let $C_1$ be any bad codebook, and let $i_1 - 1 = d(C_1)$. By Corollary 3.1, the size of $C_1|_{i_1-1} \setminus \mathcal{T}$ is at least $M^{N-i_1}$. Moreover, the size of $C_1|_{i_1-1}$ is exactly $M^{N-i_1+1}$. Let $\mathbb{B}_2$ be $\mathbb{B}_1 \setminus C_1|_{i_1-1}$. Let $C_2$ be any codebook in $\mathbb{B}_2$ and let $d(C_2) = i_2 - 1$. We now claim that the set $C_2|_{i_2-1}$ is disjoint from $C_1|_{i_1-1}$. Assume otherwise, then $C_1|_{i_1-1}$ is strictly included in $C_2|_{i_2-1}$ (recall $C_2 \notin C_1|_{i_1-1}$). This implies that $i_2 < i_1$ and $\Delta(i_2,\mathbf{x},C_1) = \Delta(i_2,\mathbf{x},C_2)$; which contradicts the minimality of $i_1 - 1 = d(C_1)$. Now as before, by Corollary 3.1, the size of $C_2|_{i_2-1} \setminus \mathcal{T}$ is at least $M^{N-i_2}$ and the size $C_2|_{i_2-1}$ is exactly $M^{N-i_2+1}$.

We continue this process iteratively, namely at step $k$, we chose a codebook $C_k \in \mathbb{B}_k$, and set $i_k - 1 = d(C_k)$. As above we have that $C_k|_{i_k-1}$ is disjoint from $C_{k'}|_{i_{k'}-1}$ for any $k' \neq k$, the size of $C_k|_{i_k-1} \setminus \mathcal{T}$ is at least $M^{N-i_k}$, and the size $C_k|_{i_k-1}$ is exactly $M^{N-i_k+1}$. We define $\mathbb{B}_{k+1}$ to be $\mathbb{B}_k \setminus C_k|_{i_k-1}$. We continue this process until $\mathbb{B}$ is entirely covered. Let $k^*$ be the last step of our procedure (*i.e.* $\mathbb{B}_{k^*+1} = \phi$). It is not hard to verify that $|\mathbb{B}| \leq \sum_{k=1}^{k^*} M^{N-i_k+1} = M \sum_{k=1}^{k^*} M^{N-i_k} \leq M |\cup_{k=1}^{k^*} C_k|_{i_k-1} \setminus \mathcal{T}| \leq M|\mathcal{T}^c|$, which concludes our assertion. ∎

Now combining the results of Lemma 3.2, 3.4 and 3.5; and setting $\lambda$ of Lemma 3.2 to be equal to $n^2$ we obtain the assertion stated in Theorem 2. In the above, by our setting of parameters, notice that $\sqrt{\lambda N(\ell+3)^2} \leq 2^{(H(p)+2R-1)n}$ (here we use the fact that $\gamma \in \left(\frac{2+H(p)}{3}, 1\right]$). The lower bound of Theorem 1 now follows easily from Theorem 2 and Lemma 2.2, full proof is given in the Appendix.

## IV. CONCLUSION

In this work we define and study the capacity of $\mathcal{W}_{p,\gamma}$ (the family of all binary $\gamma$-oblivious $p$-channels). Such families of channels arise when considering communication in an adversarial jamming model in which the jammer $Z$ is limited in resources. We limit the jammer by both a power constraint and by the restriction to impose its errors based only on a small number of possible error distributions. For $\gamma = 1$ such families are closely related to AVC's with state constraints, and it has been shown in [4], [5] that $C(\mathcal{W}_{p,1}) = 1 - H(p)$.

We show for $p < 1/2$ and $\gamma \in \left(\frac{2+H(p)}{3}, 1\right]$ that $C(\mathcal{W}_{p,\gamma})$ is at least $\gamma - H(p)$. For $\gamma = 1$ our contribution is in our new proof technique. Roughly speaking, our proof is of combinatorial nature, is based on a relatively new "strong concentration inequality" of [17], and differs substantially from the proof presented in [4], [5]. For $\gamma \in (0,1)$ this work initiates the study of $\gamma$-oblivious channels.

## ACKNOWLEDGMENTS

I would like to thank Sidharth Jaggi for several helpful discussions and comments on the oblivious channel model. Research supported in part by NSF grant CCF-0346991.

## APPENDIX

### A. Maximum error and Random coding

For a channel $W$ and a code $\mathcal{C}$ define the maximum error $e = e(\mathcal{C}, W) = \max_i e(i)$. For a family of channels $\mathcal{W}$, let $\mathcal{C}^m(\mathcal{W})$ be the capacity of $\mathcal{W}$ with respect to $e$. In this work we did not address $\mathcal{C}^m(\mathcal{W}_{p,\gamma})$ for $\gamma \in [0, 1]$ as it holds that $\mathcal{C}^m(\mathcal{W}_{p,\gamma}) = \mathcal{C}^m(\mathcal{W}_p)$. This follows directly by our definitions.

Let $\mathcal{C}^*$ be a distribution over $[n, N]$ codes. $\mathcal{C}^*$ is said to allow communication over $\mathcal{W}_{p,\gamma}$ with maximum error $\varepsilon$ if for each $W \in \mathcal{W}_{p,\gamma}$ the expected error $\mathbb{E}[e(\mathcal{C}, W)]$ is at most $\varepsilon$ (here the expectation is over $\mathcal{C}^*$). The random capacity $\mathcal{C}^r(\mathcal{W}_{p,\gamma})$ is now defined analogously to the deterministic capacity used throughout the paper. In [9], [11] it is shown that $\mathcal{C}^r(\mathcal{W}_{p,0}) = 1 - H(p)$. Let $\pi$ be the distribution over errors $\mathbf{e} \in \{0, 1\}^n$ in which $\Pr[\pi = \mathbf{e}] = p^{\|e\|}(1-p)^{n-\|e\|}$. Let $\pi'$ be $\pi$ restricted to errors $\mathbf{e}$ with Hamming weight less than or equal to $pn$. Let $W_{\pi'} \in \mathcal{W}_{p,1}$ be the channel in which $W_{\pi'}(\cdot|\mathbf{x}) = \pi'$ for all $\mathbf{x} \in \{0, 1\}^n$. It now holds that $\mathcal{C}^r(\mathcal{W}_{p,1}) \leq \mathcal{C}^r(W_{\pi'}) = \mathcal{C}^m(W_{\pi'}) \leq 1 - H(p)$ (the last inequality is proven in Section C of this Appendix). We conclude, that $\mathcal{C}^r(\mathcal{W}_{p,\gamma}) = 1 - H(p)$, for $\gamma \in [0, 1]$.

### B. Average vs. maximum error in $\mathcal{W}_p$

As above, for a channel $W$ and a code $\mathcal{C}$ define the maximum error $e = \max_i e(i)$. For a family of channels $\mathcal{W}$, let $\mathcal{C}^m(\mathcal{W})$ be the capacity of $\mathcal{W}$ with respect to $e$, and $\mathcal{C}^a(\mathcal{W})$ be the capacity of $\mathcal{W}$ with respect to $\bar{e}$.

*Lemma 1.1:* $\mathcal{C}^a(\mathcal{W}_p) = \mathcal{C}^m(\mathcal{W}_p)$.

*Proof:* It is clear that $\mathcal{C}^a(\mathcal{W}_p) \geq \mathcal{C}^m(\mathcal{W}_p)$ thus we prove the missing inequality. For a given $0 < \varepsilon < 1/4$ assume the existence of an $[n, N]$ code $\mathcal{C} = (C, \phi)$ that allows communication over $\mathcal{W}_p$ with $\bar{e} \leq \varepsilon$. Let $C = \{\mathbf{x}_1, \ldots, \mathbf{x}_N\}$. Let $N = 2^{Rn}$. We will show the existence of a subset $C'$ of $C$ of size at least $2^{Rn-1}$ s.t. using $\mathcal{C}' = (C', \phi')$ on $\mathcal{W}_p$ we obtain $e = 0$ (here $\phi'$ is the Nearest Neighbor decoder). This is enough to prove our assertion.

Consider the following graph $G$ with vertex set $C$ and an edge between $\mathbf{x}_i$ and $\mathbf{x}_j$ iff $\|\mathbf{x}_i \oplus \mathbf{x}_j\| \leq 2pn$. Let $M$ be a maximal matching in $G$, namely a maximal set of edges $M$ such that every vertex in $G$ is adjacent to at most a single edge in $M$. Consider the subgraph $G_M$ of $G$ in which we include only edges in the matching $M$. Let $I_M$ be the set of vertices in $G_M$ with no adjacent edges. $I_M$ is an independent set in $G$ (and also in $G_M$). In other words, the codebook consisting of codewords in $I_M$ has minimum distance $2pn + 1$ and thus when used with the Nearest Neighbor decoder $\phi'$ on $\mathcal{W}_p$ will have error $e = 0$. It is left to show that $|I_M|$ is large. Let $W(\mathbf{y}|\mathbf{x})$ be the following channel: 1) for codewords $\mathbf{x}_i$ with a corresponding codeword $\mathbf{x}_j$ s.t. the edge $(\mathbf{x}_i, \mathbf{x}_j)$ is in $M$, set $W(\mathbf{y}|\mathbf{x}_i) = 1$ where $\mathbf{y}$ is the center of the minimum radius ball in $\{0, 1\}^n$ including $\mathbf{x}_i$ and $\mathbf{x}_j$; 2) for the remaining $\mathbf{x} \in \{0, 1\}^n$ set $W(\mathbf{x}|\mathbf{x}) = 1$. Notice that $W \in \mathcal{W}_p$. It now follows that the average decoding error of $\mathcal{C}$ when communicating on $W$ is $\frac{|C|-|I_m|}{2|C|} \leq \varepsilon$. This implies that $|I_M| \geq (1-2\varepsilon)|C| \geq 2^{Rn-1}$. ∎

### C. Upper bound on $\mathcal{C}(\mathcal{W}_{p,1})$

Let $\pi$ be the distribution over errors $\mathbf{e} \in \{0, 1\}^n$ in which $\Pr[\pi = \mathbf{e}] = p^{\|e\|}(1-p)^{n-\|e\|}$. Let $\pi'$ be $\pi$ restricted to errors $\mathbf{e}$ with Hamming weight less than or equal to $pn$. Let $W_\pi$ be the channel in which $W_\pi(\cdot|\mathbf{x}) = \pi$ for all $\mathbf{x} \in \{0, 1\}^n$. Let $W_{\pi'}$ be the channel in which $W_{\pi'}(\cdot|\mathbf{x}) = \pi'$ for all $\mathbf{x} \in \{0, 1\}^n$. Notice that $W_{\pi'} \in \mathcal{W}_{p,1}$ and that $W_\pi = W_{BSC_p}$. We now show that $\mathcal{C}(W_{\pi'}) \leq 1 - H(p)$ (this will suffice to prove our assertion). Assume otherwise, namely that for $R > 1 - H(p)$, $\varepsilon < 1/4$ and sufficiently large $n$ there exists $[n, 2^{Rn}]$ codes $\mathcal{C}$ which allow communication over $W_{\pi'}$ within error $\varepsilon$. This implies that $\mathcal{C}$ allows communication over $W_\pi = W_{BSC_p}$ within constant error bounded away from 1. This contradicts a fundamental result on the $\varepsilon$-capacity of $W_{BSC_p}$ [14].

### D. Attempt for an alternative definition for obliviousness

An alternative definition to $\gamma$-oblivious channels $W$ is $I = \max_X I(X; Z) \leq (1-\gamma)n$. Here $X$ represents any distribution over codewords transmitted, and $Z$ denotes the error imposed by the channel. The random variables $X$ and $Z$ are jointly distributed according to $\Pr[X = \mathbf{x}, Z = \mathbf{e}] = W(\mathbf{e}|\mathbf{x})$. There are various connections between the suggested definition and the original one given in Definition 1.1. Namely, it is not hard to verify that if a channel $W$ is $\gamma$-oblivious by Definition 1.1 then it is $\gamma$-oblivious by the above definition. The other direction holds for $\gamma = 0$ or 1 but is not necessarily true for $\gamma \in (0, 1)$. For example, consider a channel $W(\mathbf{e}|\mathbf{x})$ defined by a set of errors $\{\mathbf{e_x}\}$ (each of Hamming weight at most $pn$) indexed by $\mathbf{x} \in \{0, 1\}^n$: $W(\mathbf{e_x}|\mathbf{x}) = \varepsilon + \alpha$, otherwise, for $\mathbf{e} \neq \mathbf{e_x}$ of weight at most $pn$, $W(\mathbf{e}|\mathbf{x}) = \alpha$. Here $\alpha$ is $(1-\varepsilon)/Vol(pn)$ where $Vol(pn)$ is the size of a Hamming ball of radius $pn$ in $\{0, 1\}^n$. Consider the family of channels $\mathcal{W}$ consisting of all such channels $W$. This family is $1-\varepsilon$ oblivious by the suggested definition and only $(1-H(p))$ - oblivious by Definition 1.1. It is not hard to verify that the capacity of $\mathcal{W}$ is that of $\mathcal{W}_p$. This implies a discontinuity in the capacity of $\gamma$-oblivious $p$-channels when using the suggested definition at the point $\gamma = 1$.

### E. Linear Codes

*Lemma 1.2:* Let $C$ be any $[n, 2^{Rn}]$ linear codebook. Let $\gamma \in [0, 1]$. There exists a decoder $\phi$ such that $C, \phi$ allow communication over $\gamma$-oblivious $p$-channels within error less than $1/2$ iff $C$ has minimum distance of value at least $2pn+1$.

*Proof:* Let $C$ be any codebook with minimum distance of value at least $2pn+1$. Let $\phi$ be the Nearest Neighbor decoder. Then for every $p$-channel $W$ it holds that $\bar{e} = \frac{1}{N}\sum_{i=1}^{N} e(i) = 0$, implying that $C$ allows communication over $\gamma$-oblivious $p$-channels with error 0.

Let $C = \{\mathbf{x}_1, \ldots, \mathbf{x}_{2^{Rn}}\}$ be an $[n, 2^{Rn}]$ linear codebook with minimum distance less than $2pn + 1$. Let $\phi$ be any decoder. By the linearity of $C$ this implies the existence of a codeword $\mathbf{x}^*$ of weight at most $2pn$ (where $\mathbf{x}^* \neq \mathbf{0}$). Let $\mathbf{e}_1$ be any error in $\{0,1\}^n$ of Hamming weight at most $pn$ such that $\mathbf{x}^* \in \mathcal{B}(pn, \mathbf{e}_1)$. Let $\mathbf{e}_2$ be $\mathbf{x}^* \oplus \mathbf{e}_1$. Notice that $\mathbf{e}_2$ is of Hamming weight at most $pn$. Notice also that for any codeword $\mathbf{x}$ it holds that $\mathbf{x} \oplus \mathbf{e}_1 = (\mathbf{x} \oplus \mathbf{x}^*) \oplus \mathbf{e}_2 = \mathbf{x}' \oplus \mathbf{e}_2$ (here $\mathbf{x}' = \mathbf{x} \oplus \mathbf{x}^*$ is a codeword of $C$). Consider the set $A_1 = \{\mathbf{x}_i | \phi(\mathbf{x}_i \oplus \mathbf{e}_1) = i\}$ and $A_2 = \{\mathbf{x}_i | \mathbf{x}_i \oplus \mathbf{x}^* = \mathbf{x}_j$ and $\phi((\mathbf{x}_i \oplus \mathbf{x}^*) \oplus \mathbf{e}_2) = j\}$. The sets $A_1$ and $A_2$ are disjoint. Thus at least one of the sets is of size most $2^{Rn}/2$, say $A_1$ (a similar proof can be given for $A_2$). Let $W$ be the deterministic 1-oblivious $p$-channel for which $\forall \mathbf{x}\ W(\mathbf{e}_1|\mathbf{x}) = 1$. We conclude that $\bar{e} = \frac{1}{N}\sum_{i=1}^{N} e(i) \geq 1/2$, implying that $C$ does not allow communication over 1-oblivious $p$-channels within error less than $1/2$. As $W$ is also a $\gamma$-oblivious channel for any $\gamma \in [0,1]$ we conclude our assumption. ∎

### F. List decodability of random codes

*Lemma 1.3:* Let $R \in (0, 1)$. Let $c$ be a sufficiently large universal constant. Let $\ell = \max(Vol(pn)2^{-n+Rn+1}, cn^2)$. Let $C$ be a random codebook in $\Omega[n, 2^{Rn}]$. With probability at least $1 - e^{-\ell/6}2^n$, $C$ is $[\ell, p]$ list decodable.

*Proof:* Let $\mathcal{B}$ be any ball of radius $pn$ in $\{0,1\}^n$. The expected number of points in the intersection of $C$ and $\mathcal{B}$ is $E = Vol(pn)2^{-n+Rn}$. Let $\ell = \max(2E, cn^2)$. The probability, for a specific ball $\mathcal{B}$ of radius $pn$, that $|C \cap \mathcal{B}|$ is less than $\ell$ (which is at least twice its expectation) is at least $1 - e^{-\ell/6}$. For $\ell = 2E$ this follows by applying the Chernoff bound [8]. For $\ell = cn^2 > 2E$ this follows by studying the probability that $|C \cap \mathcal{B}'| \leq \ell$ for a larger subset $\mathcal{B}'$ including $\mathcal{B}$. Thus the probability that this holds for every ball of radius $pn$ in $\{0,1\}^n$ is at least $1 - e^{-\ell/6}2^n$. ∎

### G. Proof of Theorem 1

Let $p \in [0, 1/2)$. Let $\gamma \in \left(\frac{2+H(p)}{3}, 1\right]$. Let $\varepsilon > 0$ and $\delta > 0$ be any sufficiently small constants. Let $R = \gamma - H(p) - \delta$. We show that for sufficiently large $n$ there exist $[n, 2^{Rn}]$ codes $\mathcal{C}$ which allow communication over $\gamma$-oblivious $p$-channels with error $\varepsilon$. The decoder $\phi$ used is the Nearest Neighbor decoder. By Lemma 2.2 it suffices to show the existence of codebooks $C$ for which $|A_\mathbf{e}(C)|$ is smaller than $\varepsilon 2^{(R-(1-\gamma))n}$ for every $\mathbf{e} \in \mathcal{B}(pn, \mathbf{0})$. Let $C$ be a random codebook in $\Omega[n, 2^{Rn}]$. The probability that $|A_\mathbf{e}(C)|$ is greater than $2^{(H(p)+2R-1)n+1}$ for a specific error $\mathbf{e} \in \mathcal{B}(pn, \mathbf{0})$ is at most $2^{-2n}$. This follows by Theorem 2 and Lemma 3.1. By our setting of parameters $2^{(H(p)+2R-1)n+1} \leq \varepsilon 2^{(R-(1-\gamma))n}$. Now, applying the union bound over all errors $\mathbf{e} \in \mathcal{B}(pn, \mathbf{0})$ we conclude that the probability that $|A_\mathbf{e}(C)|$ is greater than $\varepsilon 2^{(R-(1-\gamma))n}$ for any error $\mathbf{e} \in \mathcal{B}(pn, \mathbf{0})$ is at most $2^{-2n}Vol(pn) < 1$. This implies the existence of an $[n, 2^{Rn}]$ code as asserted in Theorem 1.